\documentclass[prd,onecolumn,floats,floatfix,apsrev,amsmath,amsfonts,nofootinbib,superscriptaddress]{revtex4-2}

\usepackage{color}
\usepackage{graphicx}
\usepackage{dcolumn}
\usepackage{natbib}
\usepackage{hyperref}
\usepackage{hypernat}
\usepackage{natbib}
\usepackage{soul}
\usepackage{amssymb}

\newcommand{\dd}{{\rm d}}

\newcommand{\p}{\partial}

\newcommand{\be}{\begin{equation}}
\newcommand{\en}{\end{equation}}

\newcommand{\paren}[1]{\left({#1}\right)}

\newcommand{\I}{\mathrm{I}}
\newcommand{\E}{\mathrm{E}}
\newcommand{\ex}{\mathrm{ex}}

\begin{document}
\title{Entropy and stability of an extremally charged Einstein-Born-Infeld thin shell }

\author{Ernesto~F.~Eiroa}
\email{eiroa@iafe.uba.ar}
\author{Griselda Figueroa-Aguirre}
\email{gfigueroa@iafe.uba.ar}
\affiliation{Instituto de Astronom\'{i}a y F\'{i}sica del Espacio (IAFE, CONICET-UBA), Ciudad Universitaria, 1428, Buenos Aires, Argentina.}

\author{Miguel~L.~Pe\~{n}afiel}
\email{miguelpenafiel@upb.edu}
\affiliation{Departamento de F\'{i}sica Te\'{o}rica, Instituto de F\'{i}sica, Universidade do Estado do Rio de Janeiro, Rua S\~{a}o Francisco Xavier 524, Maracan\~{a}, CEP 20550-013, Rio de Janeiro, RJ, Brazil.}
\affiliation{FIA - Facultad de Ingeniería y Arquitectura, Universidad Privada Boliviana, Camino Achocalla Km 3.5, La Paz, Bolivia.}


\begin{abstract}
Spacetimes with a thin shell offer a framework where both the dynamical and the thermodynamical stability of the matter comprising the shell can be consistently studied. In the present work, we consider the dynamical and the thermodynamical stability of a spherical thin shell in Einstein gravity coupled to Born-Infeld electrodynamics. For our construction, we adopt the extremally charged solution of the theory, which gives a closed analytic form for the horizon location that allows for a clear derivation of the corresponding physical quantities of interest. Under this scenario, the dynamical stability conditions under radial perturbations are readily obtained in terms of an effective potential. The equilibrium thermodynamics for such a shell is presented. We find that, despite a non-zero pressure at the shell (unlike the extremally charged Reissner-Nordstr\"{o}m counterpart), its entropy is solely characterized as a function of the gravitational radius. We propose a physically suitable \emph{ansatz} for the relevant equations of state in order to obtain a closed expression for the entropy density of the shell. We find that the thermodynamical stability conditions reduce to a single inequality related to exchanges of the charge at the shell, which determines the domain where both dynamical and thermodynamical stable configurations exist.

\end{abstract}

\maketitle


\section{Introduction}

General Relativity (GR) is a highly successful theory for the gravitational field that, among its most remarkable results, one can mention the predictions of astrophysical black holes and gravitational waves. Nowadays, both of them have been observed by current facilities and the measurements coincide with GR predictions (among several other alternative theories) \cite{Will2014,Abbott2016,Akiyama2019}. However, despite the enormous success of the theory, GR still has some puzzling behaviors that need to be better understood, with black hole entropy (thermodynamics) being a distinctive example.

Following the formulation of the laws of black hole mechanics \cite{Bardeen1973}, the seminal works of Bekenstein \cite{Bekenstein1972,Bekenstein1973,Bekenstein1974} and Hawking \cite{Hawking1975} led to the understanding that: (i) a black hole has an entropy, (ii) the entropy is linear with the area of the horizon and (iii) a black hole radiates (it can evaporate).  It is particularly important to notice that in the black hole case, the entropy scales with the area of the horizon rather than with some given volume as in other physical scenarios. These latter results come from a framework in which quantum fields are studied in a curved spacetime and, a complete understanding of black hole entropy will be only attainable once a full theory of quantum gravity is available. Yet, within the classical domain, it is possible the study of the thermodynamics of some shell of matter which eventually collapses. Once a classical thermodynamical analysis for the matter at the shell is performed, its macroscopic thermodynamics can be analyzed, potentially leading to the understanding of the emergence and the physical meaning of the entropy of the black hole \cite{Lemos2015,Lemos2015a,Fernandes2022}.

A self-gravitating thin shell of matter can be modeled by joining two spacetime regions across a hypersurface where matter is present. The usual prescription for building a spacetime having a thin shell is given by the Darmois-Israel formalism which provides the junction conditions for the induced metric and the extrinsic curvature across the shell \cite{Israel1966}.  Once a thin shell is constructed, it is possible to assess if the given configuration will be stable upon perturbations preserving the symmetry \cite{Ishak2002,LeMaitre2019,Mazharimousavi2017}. It is also of interest the study of the thermodynamical stability of the matter comprising the shell, which can be regarded as ensuring that it will not suffer a phase transition \cite{Martinez1996}. Both of these stability criteria are conceptually independent and, \emph{a priori}, nothing guarantees that a given thin shell can always fulfill both for the same configuration. As such, one may regard a solution to be \emph{completely stable} when both stability conditions are satisfied \cite{Bergliaffa2020,Reyes2022,Eiroa2024}.

A natural branch of the many possibilities for building a spacetime with a thin shell and studying its stability properties is to consider a shell in GR comprised of charged matter that obeys a certain extension of Maxwell electrodynamics. A widely studied example is nonlinear electrodynamics, which emerges as a generalization of Maxwell electrodynamics where the Lagrangian density can be taken as being an arbitrary function of the two Lorentz scalars constructed from the Faraday tensor $F_{\mu\nu}$ and its dual $\widetilde{F}_{\mu\nu}$ \cite{Plebaǹski1970}. A particularly interesting and fully examined case of nonlinear electrodynamics is Born-Infeld (BI) theory \cite{Born1934}, which is described by the Lagrangian density
\be
\mathcal{L}=\frac{1}{b^2}\paren{1-\sqrt{1+F b^2-\frac{G^2b^4}{4}}}\ ,
\en
where $F\equiv F_{\mu\nu}F^{\mu\nu}/2$, $G\equiv\widetilde{F}_{\mu\nu}F^{\mu\nu}/2$ are the two invariants and $b$ is the inverse of a \emph{maximum field} parameter, which has dimensions of length. Naturally, this theory offers regular results for the electric field and the energy of a point particle in the classical level and it was originally introduced by Born and Infeld as a candidate towards a quantum theory of electrodynamics. BI theory has gained a renewed interest on the theoretical side, since it emerges in the low-energy regime of string theory \cite{Fradkin1985} and, in the classical regime, has been widely studied in its coupling to GR leading to the Einstein-Born-Infeld (EBI) black hole and the geon (EBIon) solutions \cite{GarciaD.1984,Demianski1986}. Many features of the EBI black hole solution have been assessed, including its horizon structure, thermodynamics \cite{Chemissany2008,Falciano2021}, geodesic motion \cite{Linares2015}, and its scattering properties \cite{Sanchez2018}, among others \cite{Breton2003,Cai2004,Eiroa2006,Gunasekaran2012}.

The dynamical stability of the EBI thin shells has been studied in the context of gravitation in $(2+1)$ dimensions \cite{Eiroa2013}, $(3+1)$ dimensions \cite{Eiroa2011,Eiroa2012}, and thin-shell wormholes in $(3+1)$ dimensions \cite{Richarte2009,Figueroa2012}. Once the EBI black hole possesses a rich phase structure for the possible horizon configurations, due to the introduction of the parameter $b$, there is a larger set of dynamical stability configurations for an EBI thin shell, compared to the Maxwell case. Regarding the thermodynamical (and complete) stability of such shells, the $(2+1)$-dimensional extremally charged EBI thin shell has been addressed in \cite{OlmosCayo2025}, where it was found that all thermodynamical stability regions are contained within the dynamical stability regions.

In the present article, we extend the results of the complete stability of an extremally charged EBI thin shell to the spherically symmetric $\paren{3+1}$-dimensional scenario. This is mainly motivated by two reasons: (i) in the EBI solution, the extremal horizon has a closed analytic form, while the horizon locations for the sub-extremal case cannot, in general, be determined analytically, and (ii) it is of interest to analyze the thermodynamics of such extremally charged shell since there is a vivid argument in the community with respect to the value of the extremal black hole entropy \cite{Hawking1995,Ghosh1997,Hod2000,Carroll2009,Lemos2016}. This work is organized as follows. In Section \ref{sec:EBIst} we review the main properties of the EBI spacetime, emphasizing the characteristics of this solution when taking the extremal limit. In Section \ref{sec:ts-st} we construct a spacetime with a thin shell by joining two regions corresponding to an inner vacuum and an outer extremal EBI solution. In Section \ref{sec:dyn-stab}, we study the dynamical stability of this shell under radial perturbations finding, with the adoption of a linear equation of state, the stability region for the parameters that characterize the thin shell. In Section \ref{sec:therm} we study the thermodynamics for the extremally charged EBI shell, obtaining the equations of state for the inverse temperature, the pressure, and the electrostatic potential that allow us to derive the resulting entropy of the shell. With an expression for the entropy at hand, and considering a power law equation of state for the inverse temperature, in Section \ref{sec:th-stab} we evaluate the thermodynamical stability conditions that ensure that any exchange of matter within the shell will not lead to a phase transition. We also present the region where both dynamical and thermodynamical stability coexist in a suitable parameter space. Finally, in Section \ref{sec:conclusions} we discuss our results and present alternatives for future research. Throughout this work we use geometrized units in which $c=G=1$ and we adopt the \emph{mostly plus} convention for the Minkowski metric, i.e., $\eta_{\mu\nu}=\text{diag}(-1,+1,+1,+1)$.


\section{Einstein-Born-Infeld solution} \label{sec:EBIst}

\begin{figure}[t]
\centering
\includegraphics[width=0.42\textwidth]{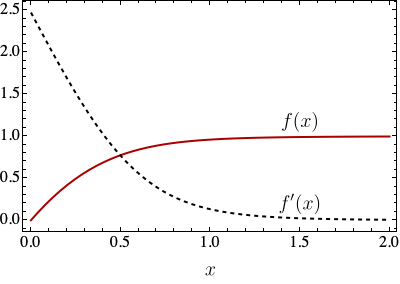}
\caption{Plot of the function $f(x)$ defined in Eq. \eqref{eq:fx} and its derivative $f'(x)$.}
\label{fig:fx}
\end{figure}

We start from the EBI spacetime, which is a solution of Einstein equations coupled to BI electrodynamics, described by the static, spherically symmetric line element of the form
\be \label{eq:EBIst}
\dd s^2=-\paren{1-\frac{2m}{r}+\frac{Q^2}{r^2}f\paren{\frac{r}{\sqrt{Qb}}}}\dd t^2+ \paren{1-\frac{2m}{r}+\frac{Q^2}{r^2}f\paren{\frac{r}{\sqrt{Qb}}}}^{-1}\dd r^2+r^2\dd\Omega^2\ ,
\en
where $m$ is the mass, $Q$ is the electric charge, and $\dd\Omega^2\equiv \dd\theta^2+\sin^2\theta\dd\varphi^2$ is the metric on a unit two-sphere. The function $f(x)$ is defined by
\be \label{eq:fx}
f\paren{x}=\frac{2}{3}x^4\paren{1-\sqrt{1+\frac{1}{x^4}}}+\frac{4}{3}\ _2F_1\left[\frac{1}{2},\frac{1}{4};\frac{5}{4};-\frac{1}{x^4}\right]\ ,
\en
where $_2F_1\left[a,b,c;z\right]$ is the Gaussian hypergeometric function, and it is useful to write its derivative $f'(x)$ as
\be \label{eq:dfx}
\frac{df(x)}{dx}=\frac{1}{x}\left[f\paren{x}+2x^4\paren{1-\sqrt{1+\frac{1}{x^4}}}\right]\ ;
\en 
both are plotted in Fig. \ref{fig:fx}. The electromagnetic field equations for EBI spacetime read
\be
\nabla_\mu\paren{\frac{F^{\mu\nu}}{\sqrt{1+F^2b^2}}}=\nabla_\mu E^{\mu\nu}=0\ , \qquad  \nabla_\mu\widetilde{F}^{\mu\nu}=0\ ,
\en
where $E_{\mu\nu}$ is denoted as the \emph{excitation tensor} and can be decomposed in terms of the fields $\mathbf{D}$ and $\mathbf{H}$ which are nonlinear functions of $\mathbf{E}$ and $\mathbf{B}$, similarly as in the case of electromagnetic fields in material media.
The only nonzero components of the electromagnetic field $F_{\mu\nu}$ and the excitation field are
\be
F_{01}(r)=\frac{Q}{\sqrt{r^4+Q^2b^2}} \ , \qquad E_{01}(r)=\frac{Q}{r^2}\ ,
\en
respectively, from which it is possible to obtain the electrostatic potential by integrating  $\mathbf{E}=-\nabla\phi$ as
\be\label{eq:phiBI}
\phi(r )=\frac{Q}{r}\ _2F_1\paren{\frac{1}{2},\frac{1}{4};\frac{5}{4};-\frac{Q^2b^2}{r^4}}\ .
\en 

Notice that the line element in Eq. \eqref{eq:EBIst} is very similar to the form of Reissner-Nordstr\"{o}m in standard coordinates. However, the function $f(x)\in\left[0,1\right]$ appearing in Eq. \eqref{eq:EBIst} incorporates the departures from Maxwell theory and it is possible to obtain either two (or one) horizons characterized by a timelike singularity at the origin that resembles Reissner-Nordstr\"{o}m (RN) spacetime, which we denote RN-phase, or one horizon characterized by a spacelike singularity similar to Schwarzschild spacetime. For a given value of the constant $b$, when the charge is small, i.e. $0\le Q/m\le Q_d/m \equiv (36\pi b/m)^{1/3} \left( \Gamma(1/4)\right) ^{-4/3}$, the function $g_{00} (r)$ has only one zero and there is a regular event horizon. For intermediate values of charge, $Q_d/m<Q/m<Q_{\ex}/m$, it has two zeros; then, as in the case of the Reissner--Nordstr\"{o}m geometry, an inner horizon and an outer event horizon exist. When $Q/m=Q_{\ex}/m$, there is one degenerate horizon. Finally, if the values of charge are large, $Q/m>Q_{\ex}/m$, the function $g_{00}(r)$ has no zeros and a naked singularity appears. In the Reissner--Nordstr\"{o}m limit ($b\rightarrow 0$) it is easy to see that $Q_d/m=0$ and $Q_{\ex}/m=1$. 
In what follows, we will restrict to the analysis of the RN-phase, delimited by the allowed values of the parameters that give an extremal horizon. In general, the roots of $g_{00}$ cannot be obtained analytically, so either some approximation procedures \cite{Falciano2021} or numerical methods are used to obtain the horizon locations. The radius of the horizon in the extremal case, determined by the simultaneous equations $g_{00}=\p_rg_{00}=0$, reads
\be \label{eq:rex}
r_{\ex}=Q\sqrt{1-\frac{b^2}{4Q^2}}\ ; 
\en
with this closed expression, we can also express the mass as a function of the charge and $b$ as
\be\label{eq:mex}
m=\frac{1}{2}\frac{Q}{\sqrt{1-b^2/4Q^2}}\paren{1-\frac{b^2}{4Q^2}+f \paren{\sqrt{\frac{Q}{b}-\frac{b}{4Q}}}} \ .
\en
Since this work is focused on analyzing the extremally charged thin shells, we have dropped the subscript `ex' on the ADM mass and charge, i.e., $m=m_{\ex}$ and $Q=Q_{\ex}$, while we keep $r_{\ex}$ for the extremal gravitational radius. 
The former expression leads to the bounds for the mass to be 
\be
Q\ge m \ge \frac{\Gamma\paren{\frac{1}{4}}^2}{6\sqrt{2\pi}}Q\approx0.874 Q\ ,
\en
which signal that the corresponding black hole can be overcharged with respect to the RN case, where the extremal limit is $m=Q$. This is due to the fact that nonlinearities of electromagnetism act as a screening mechanism for the charge. Formally, Maxwell electrodynamics is recovered as $b\to0$ and from Eq. \eqref{eq:rex} the parameter $b$ has the bounds
\be\label{eq:boundseta}
0\le b<2Q\ .
\en
Finally, replacing Eq. \eqref{eq:mex} in Eq. \eqref{eq:EBIst}, it leads to a closed expression for the extremal EBI spacetime, which is a function of $\paren{Q,b,r}$. Alternatively, the use of Eq. \eqref{eq:rex} makes possible to express the line element as a function of $\paren{m,b,r}$ or $\paren{r_{\ex},b,r}$.


\section{Spacetime with a thin shell} \label{sec:ts-st}

Let us consider a two-dimensional timelike spherical shell of radius $R$, which we denote by $\Sigma$. The shell divides the spacetime into two regions: the internal and the external ones. The internal region $0 \le r<R$, is described by a flat spacetime geometry, while the external region $r>R$, is described by the extremal EBI solution. Therefore, the metric in both parts can be specified by
\be
\dd s^2_{(\I,\E)}=g_{\alpha\beta}^{(\I,\E)}\dd x^\alpha\dd x^{\beta}=-\psi_{\paren{\I,\E}}(r)\dd t_{(\I,\E)}^2+\frac{\dd r^2}{\psi_{(\I,\E)}(r)}+r^2\dd\Omega^2\ .
\en
where $\psi_{\paren{\I,\E}}\equiv g_{00}^{\paren{\I,\E}}$ and $(\I,\E)$ refer to the internal and the external regions, respectively. In addition, $t_{(\I,\E)}$ correspond to the internal and the external time coordinates. Then, the functions $\psi_{(\I,\E)}$ are given by
\begin{align}
\psi_{(\I)}(r)&=1\ , \label{eq:psiI} \\
\psi_{(\E)}(r)&=1-\frac{2m}{r}+\frac{Q^2}{r^2}f\paren{\frac{r}{\sqrt{Qb}}}\ ,\label{eq:psiE}
\end{align}
Since we are dealing with the extremal EBI spacetime, the ADM mass and electric charge are related to the extremal horizon radius by Eqs. \eqref{eq:rex} and \eqref{eq:mex}. In our construction we remove the region inside the horizon of the external geometry, so we take $R\ge r_{ex}$. This condition, with the help of Eq. \eqref{eq:rex}, introduces an upper bound for the extremal charge in addition to the lower bound of Eq. \eqref{eq:boundseta}, so that
\be \label{eq:boundQb}
\frac{b}{2} < Q \le \sqrt{ R^2 + \frac{b^2}{4} } \ .
\en

Finally, at the hypersurface $r=R$, the metric $h_{ab}$ on $\Sigma$ is that of a 2-sphere with an additional time coordinate, so the adequate coordinates are $y^a=\paren{\tau,\theta,\varphi}$, thus we have
\be
\dd s^2_\Sigma=h_{ab}\dd y^a\dd y^b=-\dd\tau^2+R^2\paren{\tau}\dd\Omega^2\ ,\quad r=R\ ,
\en
where $\tau$ is the proper time for an observer located at the shell. The metric $h_{ab}$ is the induced metric on the hypersurface $\Sigma$ and can be written in terms of the internal (external) metric as
\be
h_{ab}^{(\I,\E)}=g^{(\I,\E)}_{\alpha\beta}e^{\alpha}_{(\I,\E)\ a}e^{\beta}_{(\I,\E)\ b}\ ,
\en
where $e^{\alpha}_{(\I,\E)\ a}$ is the tangent vector to the hypersurface as seen from the internal and the external regions. In order to develop the thin-shell formalism for this case we should evaluate the junction conditions. The first junction condition implies that the jump in the induced metric across the hypersurface is smooth, i.e.,
\be \label{eq:firstcond}
\left[h_{ab}\right]=0\ ,
\en
which directly implies that $h_{ab}^{(\I)}=h_{ab}^{(\E)}=h_{ab}$. On each of the sides of the hypersurface, the time and radial coordinates are parametrized as $t_{(\I,\E)}=T_{(\I,\E)}\paren{\tau}$, $r_{(\I,\E)}=R_{(\I,\E)}\paren{\tau}$. Thus, the first junction condition reads
\be
-\paren{1-\frac{2m}{r}+\frac{Q^2}{r^2}f\paren{\frac{r}{\sqrt{Qb}}}}\dot{T}_{(\E)}^2+\paren{1-\frac{2m}{r}+\frac{Q^2}{r^2}f\paren{\frac{r}{\sqrt{Qb}}}}^{-1}\dot{R}_{(\E)}^2=-\dot{T}_{(\I)}^2+\dot{R}_{(\I)}^2=-1\ ,
\en
where the dot denotes the derivative with respect to $\tau$. Note that the first junction condition allows us to write $\dot{T}\paren{R,\dot{R}}$ in both the internal and external regions.

The second junction condition is related to the jump in the extrinsic curvature across the hypersurface, i.e. 
$\left[K_{ab}\right]$, where 
\be
K^{a}_{(\I,\E)\ b}=\paren{\nabla_\beta n^{\paren{\I,\E}}_\alpha}e^{\alpha}_{\paren{\I,\E}\ c}e^{\beta}_{\paren{\I,\E}\ b}h^{ca}_{\paren{\I,\E}}\ ,
\en
is the extrinsic curvature on each side of the hypersurface and $n^{\paren{\I,\E}}_\alpha$ are the unit normal vectors to the shell on each side. Note that, by construction, $e^{\alpha}_{(\I,\E)\ a}n^{(\I,\E)}_{\alpha}=0$. The second junction condition determines whether the hypersurface is a boundary surface (when $\left[K_{ab}\right]=0$) or a thin shell (when $\left[K_{ab}\right]\ne0$). By assuming that the distributional part of the Einstein tensor is related to the matter on the shell, we obtain that the jump in the extrinsic curvature is related to the surface stress-energy tensor $S_{ab}$ by the Lanczos equations
\be \label{eq:Lanczos}
8\pi S_{ab}=-\left[K_{ab}\right]+h_{ab}\left[K\right]\ ,
\en
where $K\equiv K_{ab}h^{ab}$. Now, 
\be \label{eq:Kttin}
K^{\tau }_{(\I,\E)\ \tau } = \frac{\psi'_{\paren{\I,\E}}(R)+2\ddot{R}}{2\sqrt{\psi_{\paren{\I,\E}}(R)+\dot{R}^2}}
\en
and
\be \label{eq:Kththout}
K^{\theta }_{(\I,\E)\ \theta } = K^{\varphi }_{(\I,\E)\ \varphi } = \frac{1}{R}\sqrt{\psi_{\paren{\I,\E}}(R)+\dot{R}^2}\ ,
\en
where the prime denotes the derivative with respect to the radial coordinate. In use of Eqs. \eqref{eq:Kttin} and \eqref{eq:Kththout} we can directly calculate the non-zero components of the stress-energy tensor at the shell $S_{ab}$ from Eq. \eqref{eq:Lanczos}. Then, we obtain 
\be \label{eq:Stt}
S^{\tau}_{\ \tau}=\frac{1}{4\pi R}\paren{\sqrt{\psi_{(\E)}(R)+\dot{R}^2}-\sqrt{\psi_{(\I)}(R)+\dot{R}^2}} 
\en
and
\be \label{eq:Sthth}
S^{\theta}_{\ \theta}=S^{\phi}_{\ \phi}=\frac{1}{2}S^{\tau}_{\ \tau}+\frac{1}{16\pi}\paren{\frac{2\ddot{R}+\psi'_{(\E)}(R)}{\sqrt{\psi_{(\E)}(R)+\dot{R}^2}}-\frac{2\ddot{R}+\psi'_{(\I)}(R)}{\sqrt{\psi_{(\I)}(R)+\dot{R}^2}}} \ .
\en

Let us consider that the matter in the shell is that of a perfect fluid, thus the stress-energy tensor for this matter is
\be \label{eq:perfect}
S^{a}_{\ b}=\paren{\sigma+p}u^au_b+ph^a_{\ b}\ ,
\en
where $\sigma$ is the energy density and $p$ is the pressure. Consequently, the energy density, $\sigma=-S^{\tau}_{\ \tau}$, and pressure $p=S^{\theta}_{\ \theta}=S^{\phi}_{\ \phi}$ are readily found from Eqs. \eqref{eq:Stt} and \eqref{eq:Sthth}.

For the static case with shell radius $R_0$, by replacing the explicit form of the metrics,  we can see that Eqs. \eqref{eq:Stt} and \eqref{eq:Sthth} take the form
\be \label{eq:Stt2}
S^{\tau}_{\ \tau}=\frac{1}{4\pi R_0}\paren{\sqrt{1-\frac{2m}{R_0}+\frac{Q^2}{R_0^2}f\paren{\frac{R_0}{\sqrt{Qb}}}}-1} 
\en
and
\be \label{eq:Sthth2}
S^{\theta}_{\ \theta}=S^{\phi}_{\ \phi} = \frac{1}{8\pi R_0}\paren{\sqrt{1-\frac{2m}{R_0}+\frac{Q^2}{R_0^2}f\paren{\frac{R_0}{\sqrt{Qb}}}}-1}+\frac{1}{8\pi}\frac{\frac{m}{R_0^2}-\frac{Q^2}{R_0^3}f\paren{\frac{R_0}{\sqrt{Qb}}}+\frac{Q^2}{2\sqrt{Qb} R_0^2}f'\paren{\frac{R_0}{\sqrt{Qb}}}}{\sqrt{1-\frac{2m}{R_0}+\frac{Q^2}{R_0^2}f\paren{\frac{R_0}{\sqrt{Qb}}}}}\ .
\en
It is worth noting that the above expressions possess the correct Maxwell limit when $R_0/\sqrt{Qb}\gg1$ (equivalently, $\sqrt{Qb}\to0$), as expected from the construction of BI electrodynamics. The scale $\sqrt{Qb}$ can therefore be interpreted as a characteristic nonlinearity radius separating two regimes: for $R\gg\sqrt{Qb}$, the spacetime is effectively described by the RN solution, while for $R\sim\sqrt{Qb}$ the nonlinear BI corrections become relevant and cannot be neglected.

Let us now turn to the electromagnetic description of the shell. The electromagnetic potential for each of the sides of the shell needs to be specified. In particular, the inner part of the manifold is flat and therefore the potential inside the shell will be constant. On the other hand, the outer region is described by the EBI line element and hence the electrostatic potential will be given by Eq. \eqref{eq:phiBI}. Consequently, we have that on both sides of the shell the electrostatic potential is given by
\begin{align}
\phi^{\paren{\I}}&=\frac{Q}{R}\ _2F_1\paren{\frac{1}{2},\frac{1}{4};\frac{5}{4};-\frac{Q^2b^2}{R^4}} \ , \quad r<R \ , \label{eq:potI}\\
\phi^{\paren{\E}}&=\frac{Q}{r}\ _2F_1\paren{\frac{1}{2},\frac{1}{4};\frac{5}{4};-\frac{Q^2b^2}{r^4}}\ , \quad r>R\ .\label{eq:potE}
\end{align}
Since the electrically charged matter is confined to the shell, the electromagnetic fields on both sides must satisfy the appropriate junction conditions. Following Refs. \cite{Kuchar1968,Lemos2015}, we first require the continuity of the electromagnetic four-potential projected onto the shell, i.e.,
\be \label{eq:potcond}
\left[A_a\right]=0\ ,
\en
where $A^{(\I,\E)}_a=A^{(\I,\E)}_\mu e^\mu_{(\I,\E)\ a}$. For the purely electric \emph{ansatz} adopted here, $A^{(\I,\E)}_\mu=\paren{-\phi^{(\I,\E)},0,0,0}$ and Eq. \eqref{eq:potcond} leads to
\[\phi^{(\I)}\paren{R}=\phi^{(\E)}\paren{R}\ ,\]
which is naturally satisfied by the potentials in Eqs. \eqref{eq:potI} and \eqref{eq:potE}. Consequently, the tangential components of the Faraday tensor are continous, i.e.,
\be
\left[F_{ab}\right]=0\ .
\en
The discontinuity associated with the surface charge is instead carried by the excitation tensor $E^{\mu\nu}$, defined through the field equation $\nabla_\mu E^{\mu\nu}=4\pi J^\nu$.\footnote{When considering NLED with sources, the Gauss law reads $\nabla\cdot\mathbf{D}=4\pi\rho$. Since there is a nonlinear relation between $\mathbf{D}$ and $\mathbf{E}$, the charge distribution does not directly generates the electric field but rather produces an electric displacement field, $\mathbf{D}$, that in turn gives an electric field $\mathbf{E}$.} The normal projection of the excitation tensor onto the shell satisfies
\be \label{eq:condEar}
\left[E_{ar}\right]=4\pi s_a\ ,
\en
where $E^{(\I,\E)}_{ar}\equiv E^{(\I,\E)}_{\mu\nu}e^{\mu}_{(\I,\E)\ a}n^\nu_{(\I,\E)}$ and $s_a=\sigma_eu_a$ is the surface current, while $\sigma_e$ is the charge density. Since the electrostatic potential is constant in the inner region, the corresponding excitation field vanishes there, while in the exterior region one has $E_{01}=Q/r^2$. Therefore, condition \eqref{eq:condEar} immediately gives
\be \label{eq:sigmae}
\sigma_e=\frac{Q}{4\pi R^2}
\en
for the charge density at the shell.


\section{Dynamical stability}\label{sec:dyn-stab}

For the study of the dynamical stability of spherically symmetric shells, let us consider the energy density and pressure stemming from Eqs. \eqref{eq:Stt}, \eqref{eq:Sthth}, and \eqref{eq:perfect}, which read
\begin{align} 
\sigma&= -\frac{1}{4\pi R}\left( \sqrt{\psi_{(\E)}(R)+{\dot R}^2}-\sqrt{\psi_{(\I)}(R)+{\dot R}^2}\right)\  , \label{00}\\ 
p&=-\frac{\sigma}{2}+\frac{1}{16\pi}\left( \frac{2\ddot R+\psi '_{(\E)}(R)}{\sqrt{\psi_{(\E)}(R)+{\dot R}^2}}-\frac{2\ddot R+\psi '_{(\I)}(R)}{\sqrt{\psi_{(\I)}(R)+{\dot R}^2}}\right)\  ,\label{pres}
\end{align}
respectively. The equations above, or any of them combined with the conservation equation
\begin{equation} 
\frac{d(R^2\sigma)}{d\tau}+p\frac{d(R^2)}{d\tau}=0,
\label{cons}
\end{equation}
determine the evolution of the radius $R(\tau)$. In order to make an analysis of the dynamical stability of this construction, we consider small perturbations preserving the symmetry around a static solution with radius $R_0$, having the energy density
\begin{equation} 
\sigma_0= -\frac{1}{4\pi R_0}\left( \sqrt{\psi_{(\E)}(R_0)}-\sqrt{\psi_{(\I)}(R_0)}\right)  , 
\label{000}
\end{equation}
and the pressure
\begin{equation} 
p_0=-\frac{\sigma_0}{2}+\frac{1}{16\pi}\left( \frac{\psi '_{(\E)}(R_0)}{\sqrt{\psi_{(\E)}(R_0)}}-\frac{\psi '_{(\I)}(R_0)}{\sqrt{\psi_{(\I)}(R_0)}}\right) . 
\label{pres0}
\end{equation}
We say that the matter at the shell is normal when the weak energy condition (WEC) is satisfied, i.e. both inequalities $\sigma _0 \geq 0$ and $\sigma_0 +p_0 \geq 0$ are fulfilled; otherwise it is exotic.

\begin{figure}[t]
\setkeys{Gin}{width=0.48\linewidth}
    \includegraphics{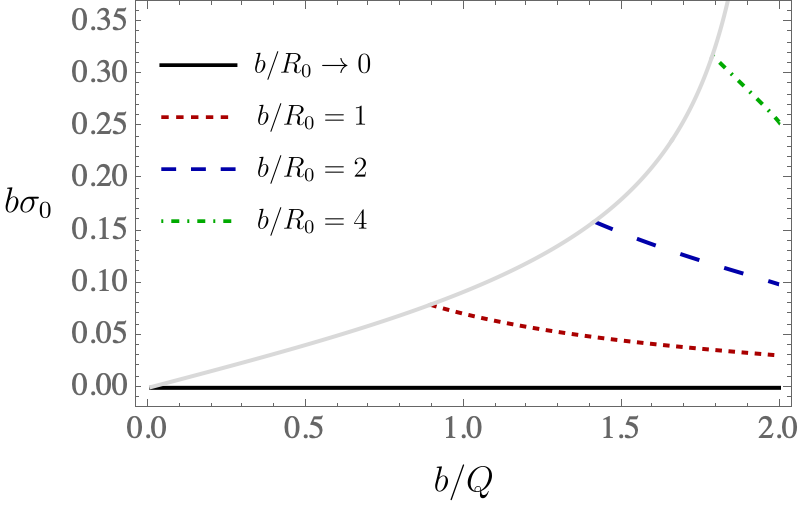} \hfill
    \includegraphics{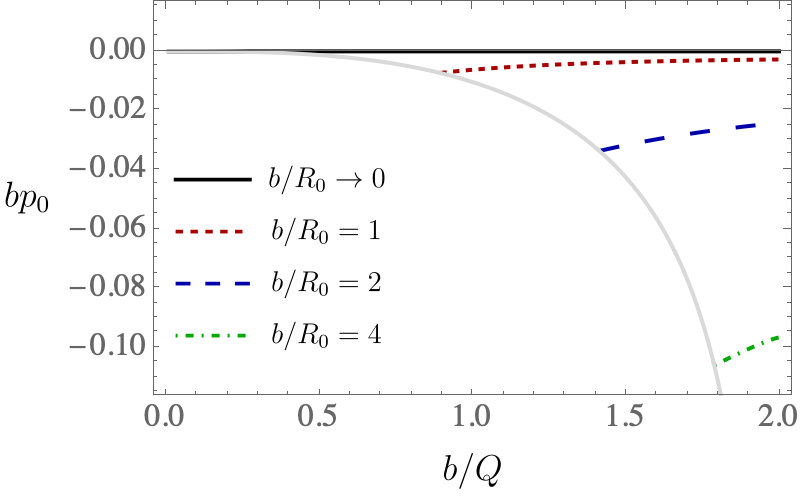}
\caption{Energy density (left panel) and pressure (right panel) as functions of $b/Q$ for the extremally charged EBI thin shell, for different values of $b/R_0$. The gray curves represent the configurations in which $R_0=r_{\ex}$, and values of $b/Q$ smaller than these critical points are unphysical in our work (see text). }
\label{fig:psigma}
\end{figure}

Provided the equation of state $p=p(\sigma)$, the conservation equation can be formally integrated 
to give $\sigma=\sigma(R)$.  After some algebraic manipulations, from Eq. (\ref{00}) we obtain
\begin{equation} 
{\dot R}^2+V(R)=0,\label{energy}
\end{equation}
where
\begin{equation} 
V(R)=\frac{\psi_{(\I)}(R)+\psi_{(\E)}(R)}{2}-\left( 2\pi R \sigma(R)\right) ^2-\left( \frac{\psi_{(\I)}(R)-\psi_{(\E)}(R)}{8\pi R \sigma(R)}\right) ^2.
\label{pot_en_func_sigma}
\end{equation}
Due to the form of Eq. (\ref{energy}), $V(R)$ can be interpreted as a potential, by analogy with the study of a particle with only one degree of freedom. The stability analysis is  based on the expansion of the potential around the static solution, given by
\begin{equation}
V(R)=V(R_0)+V'(R_0)(R-R_0)+\frac{V''(R_0)}{2}(R-R_0)^2+\mathcal{O}(R-R_0)^3\ .
\end{equation}
The first derivative of the potential reads
\begin{eqnarray}
V'(R)&=& \frac{ \psi_{(\I)}'(R)+\psi_{(\E)}'(R)}{2} -\frac{(\psi_{(\I)}(R)-\psi_{(\E)}(R)) \left(\psi_{(\I)}'(R)-\psi_{(\E)}'(R)\right)}{32 \pi ^2 R^2 \sigma (R)^2} \nonumber \\
&& -\left(2 p(R)+\sigma (R)\right) \left(\frac{(\psi_{(\I)}(R)-\psi_{(\E)}(R))^2}{32 \pi ^2 R^3 \sigma (R)^3}-8 \pi ^2 R \sigma (R)\right),
\end{eqnarray}
where we have used the conservation equation, rewritten in the form $R \sigma '(R) = -2(\sigma(R)+p(R))$. One can easily check that $V(R_0)=0$ and $V'(R_0)=0$. The second derivative of Eq. (\ref{pot_en_func_sigma}) has the form
\begin{eqnarray}
V''(R)&=& \frac{\psi_{(\I)}''(R)+\psi_{(\E)}''(R)}{2} -\frac{(\psi_{(\I)}(R)-\psi_{(\E)}(R)) \left(\psi_{(\I)}''(R)-\psi_{(\E)}''(R)\right)}{32 \pi ^2 R^2 \sigma (R)^2} \nonumber \\
&& -\frac{\left(\psi_{(\I)}'(R)-\psi_{(\E)}'(R)\right)^2}{32 \pi ^2 R^2 \sigma (R)^2} - \left(2p(R)-R \sigma '(R)\right) \frac{\left( \psi_{(\I)}(R)-\psi_{(\E)}(R)\right) \left( \psi_{(\I)}'(R)-\psi_{(\E)}'(R)\right) }{16 \pi ^2 R^3 \sigma (R)^3} \nonumber \\
&& +\left( 3\left( \sigma (R) + 2 p(R) \right) \sigma (R) + 2 R \left(\sigma (R) +3p(R) \right) \sigma '(R)-2R \sigma (R) p'(R) \right) \frac{(\psi_{(\I)}(R)-\psi_{(\E)}(R))^2}{32 \pi ^2 R^4 \sigma (R)^4} \nonumber \\
&&+8 \pi ^2 \left( \left( \sigma (R) + 2 p(R)\right) \sigma (R) + 2R \left( \sigma (R)+ p(R) \right)\sigma '(R)+ 2R\sigma (R) p'(R)\right) .
\end{eqnarray}
We adopt a linear barotropic equation of state (EoS) on the shell
\begin{equation}
p=\kappa \sigma ,
\label{equationofState}
\end{equation}
where $\kappa$ is a constant. When the parameter $\kappa $ is within the range  $0<\kappa \le 1$, it can be understood as the square of the  velocity of sound on the shell. If $\kappa >1$ this interpretation is not valid, because it would mean a speed greater than the velocity of light, implying the violation of causality. Matter with $\kappa <0$ is not common (though not impossible, e.g. $\kappa =-1$ in the Casimir effect between the plates). In this case, clearly the interpretation of $\kappa $ as the squared velocity of sound is no longer admissible. Initially we consider any real value of $\kappa $, but we will see that some restrictions apply in our study. Then, using the equation of state to write $p' =\kappa \sigma '$ and the conservation equation again, we obtain the second derivative in terms of the energy density and the pressure. For the shell radius $R_0$, by replacing the corresponding expressions of $\sigma_0$ and $p_0$, we find that
\begin{eqnarray}
V''(R_0)&=& \frac{\psi_{(\E)}(R_0)^{3/2} \left(\psi_{(\I)}'(R_0)^2-2 \psi_{(\I)}(R_0) \psi_{(\I)}''(R_0)\right)- \psi_{(\I)}(R_0)^{3/2}\left(  \psi_{(\E)}'(R_0)^2-2\psi_{(\E)}(R_0) \psi_{(\E)}''(R_0)\right) }{2 \psi_{(\I)}(R_0) \psi_{(\E)}(R_0) \left(\sqrt{\psi_{(\I)}(R_0)}-\sqrt{\psi_{(\E)}(R_0)}\right)}
 \nonumber \\
&& +\frac{(2 \kappa+1) \left(\sqrt{\psi_{(\E)}(R_0)} \left(2 \psi_{(\I)}(R_0)-R_0 \psi_{(\I)}'(R_0)\right)-\sqrt{\psi_{(\I)}(R_0)} \left(2 \psi_{(\E)}(R_0)-R_0 \psi_{(\E)}'(R_0)\right)\right)}{R_0^2 \left(\sqrt{\psi_{(\I)}(R_0)}-\sqrt{\psi_{(\E)}(R_0)}\right)}.
\end{eqnarray}

\begin{figure}[t!]
\centering
\includegraphics[width=0.42\textwidth]{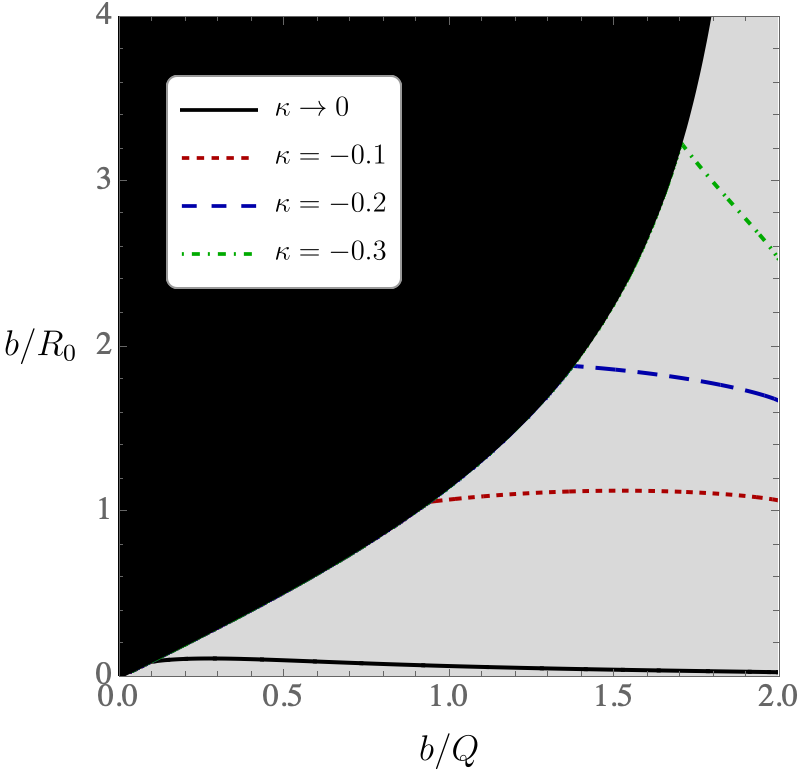}
\caption{Dynamical stability under radial perturbations of EBI thin shells in the $\paren{b/Q,b/R_0}$ plane. In the gray region the configurations with radius $R_0 > r_{\ex}$ are dynamically stable. The black zone represents shells with $R_0<r_{\ex}$, which are not of physical interest and are forbidden. The different curves correspond to some representative values of the EoS parameter $\kappa$. }
\label{fig:fig2}
\end{figure}

As stated above, in our construction we adopt the Minkowski spacetime for the internal region and the EBI spacetime for the external one. Thus, each of the regions are determined by Eqs. \eqref{eq:psiI} and \eqref{eq:psiE}, respectively.
We have to recall that the shell radius $R_0\ge r_{\ex}$, so that singularity at the center of the original outer manifold is absent; the extremal event horizon is also removed if $R_0 \neq r_{\ex}$. In this case, $R_0$, Eqs. (\ref{000}) and (\ref{pres0}) give the energy density
\begin{equation} 
\sigma_0= -\frac{1}{4\pi R_0}\left( \sqrt{\psi_{(\E)}(R_0)}-1\right)  , 
\label{bubblesigma}
\end{equation}
and the pressure
\begin{equation} 
p_0=-\frac{\sigma_0}{2}+\frac{1}{16\pi}\left( \frac{\psi'_{(\E)}(R_0)}{\sqrt{\psi_{(\E)}(R_0)}}\right)  .
\label{bubblepres}
\end{equation}
Figure \ref{fig:psigma} displays $b\sigma_0$ and $bp_0$ as functions of $b/Q$ for some representative values of $b/R_0$. The plots show that $\sigma_0>0$ and $p_0<0$ for every allowed value of $b/Q$ and, as $b/Q$ grows, the product $b\sigma_0$ decreases and $b p_0$ increases. The inequality $\sigma_0 + p_0 >0$ is always fulfilled, so the WEC is satisfied at the thin shell. Note that the opposite signs of $\sigma_0$ and $p_0$ leads to $\kappa < 0$ and the requirement that $R_0\ge r_{\ex}$ results in the bound\footnote{The value of this lower bound is obtained by replacing Eqs. \eqref{bubblesigma} and \eqref{bubblepres} in Eq. \eqref{equationofState} and taking the limits $R_0 \to r_{\ex}$ and $b/Q \to 2$.} $-1/2 < \kappa $. Therefore, the range of the EoS parameter in our construction is $-1/2 < \kappa < 0$.

The second derivative of the potential in this case reads
\begin{equation} 
V''(R_0)= \frac{ 2\psi(R_0) \psi''_{(\E)}(R_0)- \psi'_{(\E)}(R_0)^2 }{2 \psi_{(\E)}(R_0) \left(1-\sqrt{\psi_{(\E)}(R_0)}\right)} + (2 \kappa+1)\frac{ 2\sqrt{\psi_{(\E)}(R_0)} - 2 \psi_{(\E)}(R_0)+R \psi'_{(\E)}(R_0)}{R_0^2 \left(1-\sqrt{\psi_{(\E)}(R_0)}\right)}.
\label{bubbled2V}
\end{equation}
The shell is dynamically stable if the condition $V''\paren{R_0}>0$ is satisfied\footnote{Stable configurations correspond to a minimum of the potential at the equilibrium radius $R_0$. The combination of the null value of the \emph{mechanical energy} in the right hand side of Eq. \eqref{energy} and the condition $V\paren{R_0}=0$ results in a negative squared velocity in the vicinity of $R_0$, so no movement is possible. A less restrictive criterion, introduced in Ref. \cite{Visser2004}, is to consider a \emph{bounded excursion} by adding a small negative shift in the potential $V\paren{R}\to V\paren{R}-\epsilon^2$, which guarantees a positive squared velocity in the vicinity of the minimum located at $R_0$. In this case, the motion is possible but limited to a small range around $R_0$. }. The expression of the second derivative is given in terms of $\paren{R_0,b,Q}$, so it is convenient to explore the dynamically stable configurations within the plane $\paren{b/Q,b/R_0}$, with constant EoS parameter $\kappa$ represented by curves in it. This parametrization allows us to study the whole set of possible configurations for the $b/Q$ ratio, which is bounded by Eq. \eqref{eq:boundQb}.
Figure \ref{fig:fig2} displays the dynamical stability regions within the $\paren{b/Q,b/R_0}$ plane. The allowed dynamically stable configurations are depicted as gray zones, while the black ones are physically forbidden because the shell radius is smaller than the gravitational radius $r_{\ex}$. Some representative curves with constant $\kappa $ (within the possible range) are shown. For the static configurations, Eqs. \eqref{bubblesigma} and \eqref{bubblepres} determine the corresponding equilibrium values of the surface energy density $\sigma_0$ and the pressure $p_0$. Once the linear equation of state in Eq. \eqref{equationofState} is adopted, the static shell should also satisfy $p_0=\kappa\sigma_0$. The constant-$\kappa$ curves displayed in Fig. \ref{fig:fig2} thus identify the subset of static extremal shells compatible with the adopted equation of state, along which the dynamical stability condition is evaluated. As the nonlinearity parameter grows, the dynamically stable region, which coincide with the physically viable one, become larger. Thus, as BI nonlinearities become more evident, the shell possesses a larger stability domain. For instance, as $b/Q\to2$ the forbidden regions are those for which $b/R_0\ge b/r_{\ex}\approx2/\sqrt{2-b/Q}\to\infty$. Hence, in this limit, all configurations of $b/R_0$ are dynamically stable.


\section{Thermodynamics of the thin shell} \label{sec:therm}

Let us now assume that the shell is in static equilibrium\footnote{For simplicity, we omit the index ``0'' from now on.} at a radius $R$. For our purposes, it is useful to define the redshift function $k$ at the shell by
\be \label{eq:redshift}
k=\sqrt{\psi_{(\E)}(R)}=\sqrt{1-\frac{2m}{R}+\frac{Q^2}{R^2}f\paren{\frac{R}{\sqrt{Qb}}}}\ .
\en
Notice that the shell radius is necessarily larger than the gravitational radius, i.e.  $R\ge r_{\ex}$, and in the limit $R\to r_{\ex}$ we have $k=0$ for all $b$. 
The surface energy density and pressure can be written in terms of the redshift function as
\be \label{eq:sigma}
\sigma =\frac{1}{4\pi R} \paren{1-k} \ ,
\en
and
\be \label{eq:pBI1}
p=\frac{1}{16\pi R^3 k}\left[R^2\paren{1-k}^2-Q^2f\paren{\frac{R}{\sqrt{Qb}}}+\frac{Q^2R}{\sqrt{Qb}} f'\paren{\frac{R}{\sqrt{Qb}}}\right] \ ,
\en
or, using the expression for the derivative of $f(x)$ shown in Eq. \eqref{eq:dfx},
\be \label{eq:pBI2}
p=\frac{1}{16\pi R^3 k}\left[R^2\paren{1-k}^2+2\frac{R^4}{b^2}\paren{1-\sqrt{1+\frac{Q^2b^2}{R^4}}}\right] \ , 
\en
Next, let us define the material mass $M$ as the matter contained in the shell of radius $R$, therefore we have $M=4\pi R^2\sigma$. We can write $M$ in terms of the redshift function in the form
\be \label{eq:M}
M=R\paren{1-k}\ .
\en
From the above definition, it is straightforward to obtain that
\be\label{eq:dMdR}
\paren{\frac{\p M}{\p R}}_{m,Q,b}=-8\pi Rp\ .
\en
By using Eqs. \eqref{eq:redshift} and \eqref{eq:M}, we can express the ADM mass in terms of $(M, Q, R)$ as
\be\label{eq:mMRQ}
m=M-\frac{M^2}{2R}+\frac{Q^2}{2R}f\paren{\frac{R}{\sqrt{Qb}}}\ ;
\en
so the pressure can now be expressed as
\be \label{eq:pMQ}
p=\frac{1}{16\pi R^2\paren{R-M}}\left[M^2+2\frac{R^4}{b^2}\paren{1-\sqrt{1+\frac{Q^2b^2}{R^4}}}\right]\ .
\en

When the thin shell is extremally charged, the relations \eqref{eq:rex} and \eqref{eq:mex} hold, and in the Maxwell limit (as $b\to0$) we have $m=Q=M=r_{\ex}$ and 
\begin{align}
\sigma_{\text{RN}}&=\frac{m}{4\pi R^2}\ , \label{eq:sigmaRN}\\
p_{\text{RN}}&=0\ . \label{eq:pRN}
\end{align}
This situation changes when considering the extremal EBI spacetime. In such a case, the expressions for the energy density and pressure are given by Eqs. \eqref{eq:sigma} and \eqref{eq:pBI2} with the corresponding substitutions. The expressions for these quantities are quite intricate and, instead of displaying the full expressions, we analyze the leading order corrections in $b$, obtaining
\begin{align}
\sigma&=\frac{M}{4\pi R^2}\approx\frac{r_{\ex}}{4\pi R^2}+\paren{R-r_{\ex}}\paren{\frac{4R^3+3R^2r_{\ex}+2R r_{\ex}+r_{\ex}^3}{160\pi R^6r_{\ex}}}b^2+\mathcal{O}\paren{b^4}\ , \\
p&\approx\frac{M^2-r_{\ex}^2}{16\pi R^2\paren{R-M}}-\frac{\paren{R^4-r_{\ex}^4}}{64\pi R^6\paren{R-M}}b^2+\mathcal{O}\paren{b^4} \approx -\frac{\paren{R^3+2R^2r_{\ex}+3R r_{\ex}^2-4r_{\ex}^3}}{320\pi R^6}b^2+\mathcal{O}\paren{b^4}\ ,
\end{align}
respectively. Notice that as $b\to0$ the expressions \eqref{eq:sigmaRN} and \eqref{eq:pRN} are recovered for the energy density and pressure, respectively. As $b$ becomes non-negligible, the ADM mass, $m$, and the material mass, $M$, will no longer be linearly related and, more importantly, the pressure for the extremally charged shell acquires negative values. 

Let us recall that in the extremal RN case it is possible to relate the ADM mass, material mass and charge identically as $m=M=Q$, a fact that is related to the equivalence between the mass and charge densities; this will no longer be possible in the extremal EBI case. For the extremal case, the ratio between the material mass and charge is
\be \label{eq:sigmasigmae}
\frac{M}{Q} = \frac{R}{\sqrt{b^2+4 r_{\ex}^2}} \paren{2-\sqrt{\frac{4 (R-r_{\ex})}{R}+\frac{b^2+4 r_{\ex}^2}{R^2} \paren{f\paren{\frac{\sqrt{2} R}{\sqrt{b}(b^2 +4r_{\ex}^2)^{1/4}}}-\frac{R}{r_{\ex}} f\paren{\frac{\sqrt{2} r_{\ex}}{\sqrt{b}(b^2 +4r_{\ex}^2)^{1/4}}}}}} \ .
\en 
The above expression possesses two limits, namely as $R\to r_{\ex}$ and as $R\to\infty$. In particular, we have 
\begin{align}
\left.\frac{M}{Q}\right|_{R=r_{\ex}}&=\frac{2r_{\ex}}{\sqrt{b^2+4r_{\ex}^2}}\ , \\
\left.\frac{M}{Q}\right|_{R\to\infty}&=\frac{r_{\ex}}{\sqrt{b^2+4 r_{\ex}^2}}+\frac{\sqrt{b^2+4 r_{\ex}^2}}{4r_{\ex}}f\paren{\frac{\sqrt{2} r_{\ex}}{\sqrt{b}(b^2 +4r_{\ex}^2)^{1/4}}}\ ,
\end{align}
respectively. Thus, the ratio $M/Q$ has the bounds
\be\label{eq:boundsMQ}
\left.\frac{M}{Q}\right|_{R=r_{\ex}}\le\frac{M}{Q}\le\left.\frac{M}{Q}\right|_{R\to\infty}\le1\ ,
\en
where the equality in both sides is saturated only when $b=0$, i.e., in the Maxwell case. 

Assuming that the shell has a well defined temperature $T$ and an entropy $S$ which is a function of the extensive variables of the system (the internal energy/material mass of the shell, its area, and eventually other ones), the first law of thermodynamics reads
\be \label{eq:firstlaw}
T dS=dM+p\ dA+\sum_{i}d W_{i}\ ,
\en
where $d W_{i}$ are work differentials to be determined by the properties of the shell. In particular, we are interested in the case of a spherical thin shell of radius $R$ within EBI theory that is extremally charged, which is characterized by the mass $M$, the area $A$, the charge $Q$, and eventually the BI parameter $b$. For a shell with an entropy function $S=S\paren{M,A,Q}$, we can write the first law as
\be
T dS= dM+pdA-\Phi dQ\ ,
\en
where $\Phi$ is the thermodynamic electric potential. Defining the inverse temperature to be $\beta_T\equiv1/T$, the latter equation takes the form
\be \label{eq:dS1}
d S=\beta_T\paren{d M+pdA-\Phi dQ}\ .
\en
In order for $dS$ to be an exact differential, the general integrability conditions
\begin{align}
\paren{\frac{\p\beta_T}{\p R}}_{M,Q}&=8\pi R\paren{\frac{\p\beta_T p}{\p M}}_{R,Q}\ , \label{eq:intcond1a}\\
\paren{\frac{\p\beta_T}{\p Q}}_{M,R}&=-\paren{\frac{\p \beta_T\Phi}{\p M}}_{R,Q}\ , \\
\paren{\frac{\p\beta_T\Phi}{\p R}}_{M,Q}&=-8\pi R\paren{\frac{\p\beta_T p}{\p Q}}_{M,R}\ ,
\end{align}
should be satisfied.

Considering that $b$ is a constant, we have that for the shell with extremal charge $Q=Q\paren{r_{\ex}}$ and $M=M\paren{r_{\ex},R}$, Eq. \eqref{eq:intcond1a} can be written as
\be
\paren{\frac{\p\beta_T}{\p R}}_{r_{\ex}}=8\pi R \beta_T\paren{\frac{\p p}{\p M}}_{Q,R} =\beta_T\frac{b^2\paren{1-k^2}+2R^2\paren{1-\sqrt{1+\frac{b^2Q^2}{R^4}}}}{2b^2k^2R}\ ,
\en
which has the general solution 
\be\label{eq:solalpha}
\beta_T=\mathrm{a}\paren{r_{\ex}}k\paren{r_{\ex},R}\ ,
\en 
where $\mathrm{a}=\mathrm{a}\paren{r_{\ex},R\to\infty}$ is regarded as the inverse temperature of the shell if the radius were infinite. From the integrability conditions, for the electrostatic potential we obtain that  
\be
\paren{\frac{\p p}{\p Q}}_{R,M}+\frac{1}{8\pi R}\paren{\frac{\p \Phi}{\p R}}_{r_{\ex}}+\Phi\paren{\frac{\p p}{\p M}}_{R,Q}=0\ .
\en
Upon substitution of the corresponding derivatives, we arrive to
\begin{align}
k\paren{\frac{\p p}{\p Q}}_{R,M}+\frac{1}{8\pi R}\paren{\frac{\p\paren{\Phi k}}{\p R}}_{r_{\ex}}-\frac{\Phi}{8\pi R}\paren{\frac{\p k}{\p R}}_{r_{\ex}}+k\Phi\paren{\frac{\p p}{\p M}}_{R,Q}&=0 \nonumber \\
k\paren{\frac{\p p}{\p Q}}_{R,M}+\frac{1}{8\pi R}\paren{\frac{\p\paren{\Phi k}}{\p R}}_{r_{\ex}}&=0 \\ 
\sqrt{R^4+b^2Q^2}\paren{\frac{\p\paren{\Phi k}}{\p R}}_{r_{\ex}}-Q&=0\ , \label{eq:poteos}
\end{align}
where $Q=\sqrt{r_{\ex}^2+b^2/4}$ in the latter. As expected, the limit $b\to0$ returns the equation for Maxwell electrodynamics (c.f. \cite{Lemos2015}). The solution of Eq. \eqref{eq:poteos} is 
\be\label{eq:solPhi}
\Phi\paren{r_{\ex},R}=\frac{\phi\paren{r_{\ex}}-\frac{Q}{R}\ _2F_1\left[\frac{1}{2},\frac{1}{4};\frac{5}{4};-\frac{Q^2b^2}{R^4}\right]}{k}\ ,
\en
where $\phi\paren{r_{\ex}}$ can be understood as the electrostatic potential if the shell radius were infinite, while the second term is just the electrostatic potential over the shell of radius $R$. This completes the description for the equations of state from the integrability conditions.

Turning back to the entropy differential, it is useful to write it as a function of $\paren{r_{\ex},R}$. For the extremal case, the differentials of $M$ and $Q$ are
\begin{align}
dQ&=\paren{\frac{\p Q}{\p r_{\ex}}}_Rd r_{\ex} , \\
dM&=\paren{\frac{\p M}{\p r_{\ex}}}_Rd r_{\ex}+\paren{\frac{\p M}{\p R}}_{r_{\ex}}d R\ . 
\end{align}
From $M=R\paren{1-k}$ we have 
\be
\paren{\frac{\p M}{\p R}}_{m,Q,b}=\paren{1-k}-R\paren{\frac{\p k}{\p R}}_{m,Q,b}=-8\pi Rp\ ,
\en
and the entropy differential, Eq. \eqref{eq:dS1}, can now be written as
\begin{align}
dS&=\left[\beta_T\paren{\frac{\p M}{\p r_{\ex}}}_R-\beta_T\Phi\paren{\frac{\p Q}{\p r_{\ex}}}_R\right]dr_{\ex}+\beta_T p dA-8\pi R\beta_T p dR \ , \nonumber\\
dS&=\beta_T\left[\paren{\frac{\p M}{\p r_{\ex}}}_R-\Phi\paren{\frac{\p Q}{\p r_{\ex}}}_R\right]dr_{\ex}\ \label{eq:dSresult},
\end{align}
where we have used the fact that $dA=8\pi RdR$. By expressing the entropy differential in Eq. \eqref{eq:dSresult} in terms of the subsidiary extensive variables $\paren{r_{\ex},R}$ and using the relations for the extremal EBI spacetime we have been able to reduce the system to be described by a single extensive parameter $r_{\ex}$. Moreover, let us note that in the treatment of the extremally charged shell within RN spacetime (c.f. Ref. \cite{Lemos2015a}), the pressure is identically zero (which in our case is recovered in the $b\to0$ limit), and the term involving $dA$ is also identically zero. On the other hand, in our case, we have considered the general situation with a non-zero pressure and, remarkably, the term multiplying $dA$ exactly cancels the contribution stemming from the material mass and the overall entropy differential is expressed in terms of $dr_{\ex}$ alone.

\begin{figure}[t]
\includegraphics[width=0.42\textwidth]{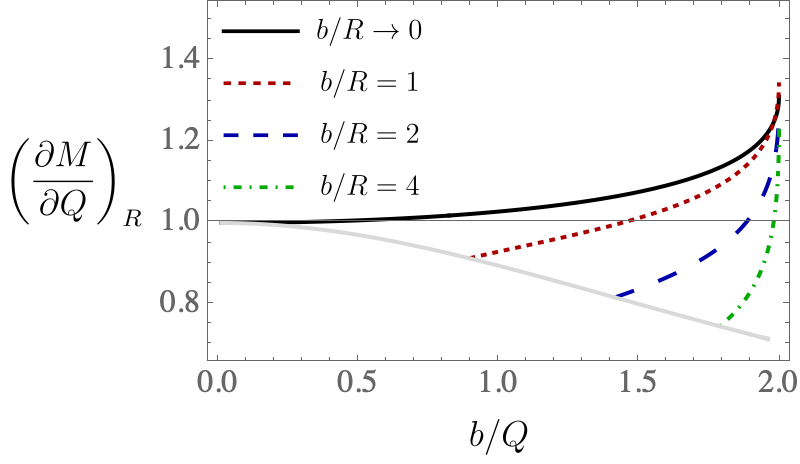}
\caption{Plot of the derivative of $M$ with respect to $Q$ for constant $R$, as a function of $b/Q$ for some values of $b/R$. The allowed values of the electrostatic potential $\Phi$ lie below each of these curves. The gray line represents the solution for this derivative when $R=r_{\ex}$. }
\label{fig:dMdQ}
\end{figure}

In order for $dS$ to be exact, from Eq. \eqref{eq:dSresult} we obtain that the integrability condition reduces to 
\be\label{eq:intcond}
\beta_T\left[\paren{\frac{\p M}{\p r_{\ex}}}_{R}-\Phi\paren{\frac{\p Q}{\p r_{\ex}}}_R\right]=s\paren{r_{\ex}}\ .
\en
Since the entropy is positive definite, we have that the integrand should be nonnegative, i.e., $s(r_{\ex})\ge0$, and the temperature is also positive definite, the term in square brackets needs to be nonnegative, thus the thermodynamic electrostatic potential need to respect the bound 
\be \label{eq:boundPhi}
\Phi\le\frac{\paren{\p M/\p r_{\ex}}_R}{\paren{\p Q/\p r_{\ex}}_R}\ ,
\en
or
\be
\Phi\le\paren{\frac{\p M}{\p Q}}_R\ .
\en
When considering the Maxwell limit we have $M=Q$ and $\Phi\le1$ (c.f. \cite{Lemos2015a}). Note that in this limit, the partial derivative is independent of $R$. For other cases it is necessary to evaluate the corresponding derivative for each value of $R$. Figure \ref{fig:dMdQ} shows $(\p M/\p Q)_R$ for some representative values of $b/R$; in order to have a positive definite entropy density the allowed values of $\Phi$ lie below each of these curves. 

Given the integrability condition in Eq. \eqref{eq:intcond}, we have that the entropy differential is 
\be \label{eq:dSBIex}
dS=s\paren{r_{\ex}}dr_{\ex}\ .
\en
Notice that the product in the left hand side of Eq. \eqref{eq:intcond} is only a function of $r_{\ex}$, despite the fact that all functions appearing on it depend on $\paren{r_{\ex},R}$. This can be verified by plugging Eqs. \eqref{eq:solalpha} and \eqref{eq:solPhi} into \eqref{eq:intcond}.

The next step is the adoption of an equation of state for the inverse temperature. The fact that the entropy of the shell of the system under consideration is solely dependent on the gravitational radius $r_{\ex}$ is truly remarkable. One may start by considering the case where the shell radius coincides with the gravitational radius and the implications towards the entropy of an extremal black hole (c.f. \cite{Lemos2015a,Lemos2016}). Nonetheless, we will study the implications of the entropy towards the thermodynamical stability analysis for some specific equations of state for the inverse temperature and the electrostatic potential. As a starting point, one may consider an equation of state for the inverse temperature of the Hawking type; however, such a choice comes out as a divergent quantity for the extremal case and therefore will be discarded when treating the extremally charged shell. A second possible equation of state for the inverse temperature is to consider a power law for the ADM mass, that is
\be \label{eq:ansatza}
\mathrm{a}=\gamma m^{\omega}\ ,
\en
where $\gamma$ is a positive constant to be determined from the matter content of the shell and $\omega$ is a free parameter that can, in principle, take any value. Inserting the charge and the ADM mass for the extremally charged shell given by Eqs. \eqref{eq:rex} and \eqref{eq:mex} into \eqref{eq:M}, we obtain the extremal material mass for the shell, which reads
\be\label{eq:93}
M=R\paren{1-\sqrt{1-\frac{2m}{R}+\frac{Q^2}{R^2}f\paren{\frac{R}{\sqrt{b Q}}}}}\ .
\en
Also, the ADM mass can be expressed as a function of $\paren{M,Q,R}$ as shown in Eq. \eqref{eq:mMRQ}.

Adopting the \emph{ansatz} \eqref{eq:ansatza} into the expression \eqref{eq:intcond} still leads to some freedom for choosing an adequate function for $\phi\paren{r_{\ex}}$ as long as the bound \eqref{eq:boundPhi} is respected. For instance, we can explore the case where
\be
\phi\paren{r_{\ex}}=\frac{Q}{2r_{\ex}}\ _2F_1\paren{\frac{1}{2},\frac{1}{4};\frac{5}{4};-\frac{Q^2b^2}{r_{\ex}^4}}\ .
\en
Then, the entropy differential for the shell given the \emph{ansatz} for the inverse temperature and electrostatic potential is
\be \label{eq:entden1}
\dd S=\frac{1}{2} \gamma  \, _2F_1\left(\frac{1}{4},\frac{1}{2};\frac{5}{4};-\frac{Q^2b^2}{ r_{\ex}^4}\right) \left(\frac{2Q^2 \, _2F_1\left(\frac{1}{4},\frac{1}{2};\frac{5}{4};-\frac{Q^2 b^2}{r_{\ex}^4}\right)}{3 r_{\ex}}+\frac{r_{\ex}}{3}\right)^{\omega }\dd r_{\ex}\ .
\en
Considering that $\omega$ is a free parameter that will allow for a complete study of the thermodynamical stability, the latter equation cannot be analytically integrated. Hence, it is not possible to find an explicit expression for the entropy of the shell. Nonetheless, since we are interested in studying the stability properties of the entropy, this will not be an obstacle. Taking into account that Eq. \eqref{eq:entden1} has the same form as Eq. \eqref{eq:dSBIex}, the right hand side can be interpreted as an \emph{entropy density} for the shell. Since the integral of $s(r_{\ex})$ gives the shell entropy and $s(r_{\ex})$ is a single valued function, then it can be understood as the derivative of $S$ with respect to $r_{\ex}$, i.e. $s(r_{\ex})=dS/d r_{\ex}$.

Finally, one may wonder whether or not the choices for the inverse temperature and electrostatic potential equations of state are correct on physical grounds. We may consider the limit where Maxwell electrodynamics is recovered from BI-NLED. The limit $b\to0$ for the entropy density leads to   
\be\label{eq:sdenRN}
dS=\frac{\gamma}{2}\paren{r_{\ex}}^\omega \dd r_{\ex}\ ,
\en
which can be easily integrated to obtain an entropy for the extremally charged RN shell that scales with its area for $\omega=1$, coinciding with previous results (c.f. \cite{Lemos2015a}).


\section{Thermodynamical and complete stability} \label{sec:th-stab}

Let us consider the case where the entropy of the shell is a function of its material mass $M$, the area $A$, and the charge $Q$, i.e., $S=S\paren{M,A,Q}$. Although the general thermodynamic state space is characterized by the independent variables $(M,A,Q)$, in the present work we restrict our attention to the extremally charged sector. Consequently, the thermodynamical stability conditions derived below describe the intrinsic stability of this extremal equilibrium configurations.

The maximum principle for the entropy implies that allowing the system for internal exchanges in each of the extensive parameters will not result in a phase transition as long as the overall entropy after the exchange is less than the initial one, i.e. the entropy should be a concave function. Such a condition can be also applied locally and, accounting for all possible exchanges, imply relations among the second derivatives of the entropy with respect to the relevant extensive parameters of the system (c.f. \cite{Callen1985,Lemos2015}). For the case under consideration, the resulting entropy is solely a function of $r_{\ex}$ and can be written in terms of $Q$ by using Eq. \eqref{eq:rex}. Since we are dealing with an entropy which only depends on one extensive parameter, the requirement of concavity for the entropy function is given by the single relation
\be
\frac{d^2S}{dQ^2}\le0\ .
\en
In turn, from the fact that the integrand in \eqref{eq:entden1} is actually $s(r_{\ex})=dS/dr_{\ex}$, this condition can be expressed in the form
\be
\frac{d S}{d r_{\ex}}\frac{\p^2r_{\ex}}{\p Q^2}+\frac{d^2S}{dr_{\ex}^2}\paren{\frac{\p r_{\ex}}{\p Q}}^2\le0\ ,
\en
or in terms of the entropy density as
\be
s\frac{\p^2r_{\ex}}{\p Q^2}+\frac{ds}{dr_{\ex}}\paren{\frac{\p r_{\ex}}{\p Q}}^2\le0\ .
\en
By computing the respective derivatives, we can see that the above condition sets an upper bound on the parameter $\omega$ given by 
\be\label{eq:omegacrit}
\omega_\text{crit}=-\frac{1}{3}-\frac{b^2/Q^2-4-\paren{b^2/Q^2+4}H}{24H^2}\ ,
\en
where we have defined 
\be
H\equiv\ _2F_1\left[\frac{1}{2},\frac{1}{4};\frac{5}{4};-\frac{16}{\paren{b^2/Q^2-4}^2}\right]\ .
\en
On the other hand, from the inverse temperature equation of state in Eq. \eqref{eq:ansatza} we have that $\omega>-1$ in order to avoid divergences in the temperature. Hence, the thermodynamically stable region corresponds to values of $\omega$ within the range
\be
-1<\omega\le\omega_\text{crit}\ .
\en

\begin{figure}[t!]
\centering
\includegraphics[width=0.9\textwidth]{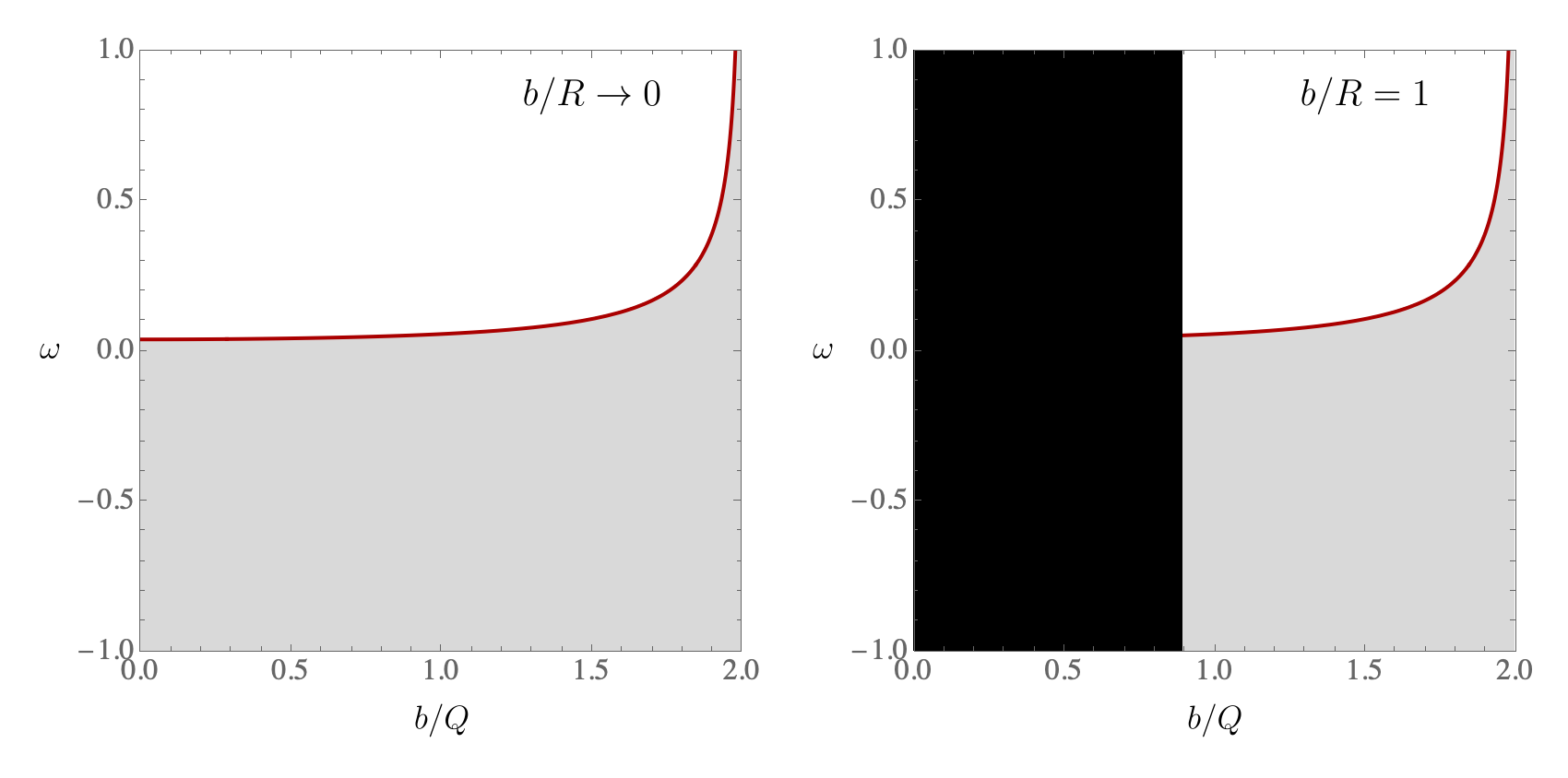}
\caption{Thermodynamical stability for the extremely charged EBI thin shell in the  $(b/Q,\omega)$ plane for two different values of $b/R$. In each plot, the continuous curve corresponds to $\omega_\text{crit}$, the gray region represents the thermodynamically stable configurations, and the black one is non-physical.}
\label{fig:thstabgridalpha}
\end{figure}

In Fig. \ref{fig:thstabgridalpha}, the thermodynamical stability for the extremally charged EBI thin shell is represented within the $\paren{b/Q,\omega}$ plane for two representative values of $b/R$. The possible values of $b/Q$ are limited by Eq. \eqref{eq:boundQb}. We can see that the stable region grows as $b/Q$ increases; in particular, the configuration is stable in the limit $b/Q\to2$ for any $\omega >-1$. Moreover, since the thermodynamical stability criterion only involves derivatives with respect to $Q$, and the entropy is not a function of $R$, shells with different radii will share the same stability properties.

\begin{figure}[t!]
\centering
\includegraphics[width=0.47\textwidth]{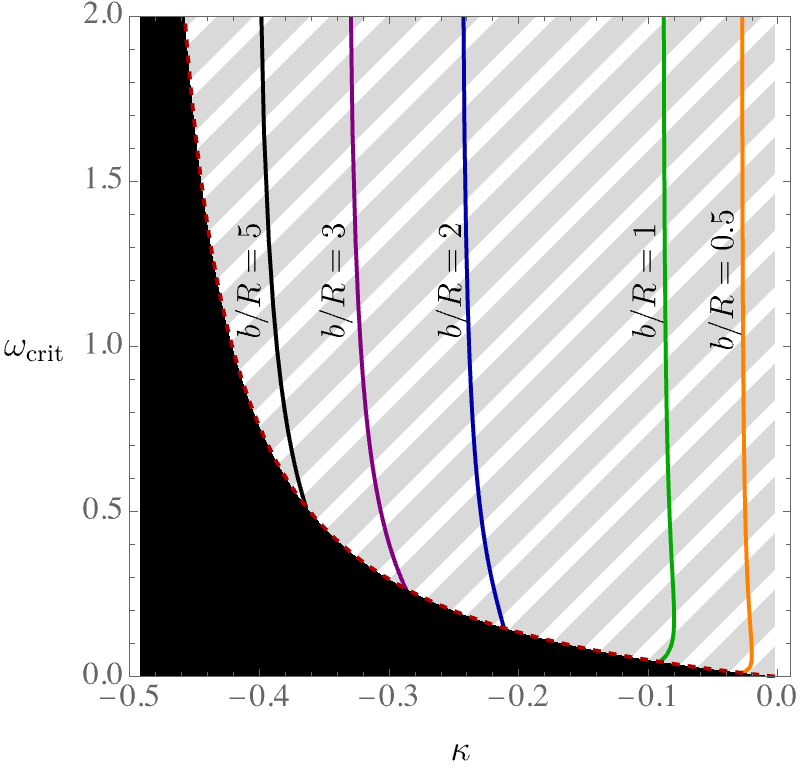}
\caption{The stripped gray region displays the possible values of $\kappa$ and $\omega_\text{crit}$. The dashed curve represents the minimum $\omega_\text{crit}$ for each $\kappa$ and the continuous curves depict some $b/R=$ constant curves in this parameter space. The black zone is unphysical.}
\label{fig:6}
\end{figure}

Let us introduce the concept of complete stability of a thin shell, defined as both dynamically and thermodynamically stable. In our study, the dynamical stability condition determines the stable region in the $\paren{b/Q,b/R}$ parameter space, in which the curves correspond to different values of $\kappa$. On the other hand, the thermodynamical stability condition gives the stable configurations in the $\paren{b/Q,\omega}$ plane, where the limiting value $\omega_\text{crit}$ is given by Eq. \eqref{eq:omegacrit}. Nonetheless, \emph{a priori}, there is no relation between $\kappa$ and $\omega$. It is not difficult to see that by obtaining $k$ from Eq. \eqref{eq:sigma} and introducing it into Eq. \eqref{eq:solalpha}, with the help of Eq. \eqref{eq:ansatza}, we can connect the inverse temperature with the surface energy density
\be
\frac{1}{T} = \gamma m^{\omega}\left(1-4\pi R \sigma \right),
\en
where $m=m(r_{\ex})$ is obtained from Eqs. \eqref{eq:rex} and \eqref{eq:mex}. Furthermore, we can relate the energy density with the pressure by using the equation of state, i.e. Eq. \eqref{equationofState}, so we can write 
\be\label{eq:invtemp}
\frac{1}{T} = \gamma m^{\omega}\left(1-\frac{\sqrt{4\pi A}}{\kappa }p\right),
\en
where $A=4\pi R^2$. The equation above corresponds to the equation of state relating $T$, $A$, and $p$ for the fluid at the shell. It is worth noticing that the dynamical stability analysis depends on $\kappa$ but not on $\omega$ while the thermodynamical stability shows the opposite behavior. These parameters are not related by any other equation except for Eq. \eqref{eq:invtemp} that determines the temperature of the shell for chosen values of $b$, $R$, and $Q$ (which give $m$, $A$, and $p$). The only way to relate them is by fixing the temperature. There is no  direct relation between $\omega$ and $R$, since thermodynamical stability of the shell does not depend on it. The dynamical stability region, displayed in Fig. \ref{fig:fig2}, shows that when the shells are physically possible, they are also dynamically stable. However, $\omega$ has the upper bound $\omega_\text{crit}$ for the thermodynamical stability, given by Eq. \eqref{eq:omegacrit}, which is solely a function of $b/Q$. Since all shells with different possible radii share the same thermodynamical stability properties, the curves with $\omega_\text{crit}=\text{const.}$ in the $\paren{b/Q,b/R}$ plane in Fig. \ref{fig:fig2} will be just vertical lines. By using Eq. \eqref{equationofState}, we can write $b/Q$ in terms of $b/R$ and $\kappa$ and consequently relate $\omega_\text{crit}$ with $b/R$ and $\kappa$, as it is displayed in Fig. \ref{fig:6}. The dashed curve in this plot represents the minimum value of $\omega_\text{crit}$ for each $\kappa$ and the possible configurations lie in the stripped gray region; some curves for representative constant values of $b/R$ are depicted by solid lines. The physical requirement that $R\ge r_\ex$ is not fulfilled in the black zone. In brief, the extremally charged thin shell is completely stable when $-1 < \omega \le \omega_\text{crit}$, with $\kappa$ and $\omega_\text{crit}$ belonging to the stripped gray region of Fig. \ref{fig:6}.

\section{Discussion and conclusions} \label{sec:conclusions}

The charged thin shells constructed in this work by gluing two regions, an inner flat geometry with an external EBI one, offer an opportunity for the study of both the dynamical and the thermodynamical stability behavior of the spherically symmetric configurations. We have considered in detail the particular case of an EBI thin shell with extremal charge, which --in contrast to the non-extremal scenario-- admits a completely analytical treatment. After introducing the study of the dynamical stability, we have developed the thermodynamics of the shell, restricting the corresponding thermodynamical stability analysis to the extremal family.

We have considered a perfect isotropic fluid at the shell with surface energy density $\sigma $ and pressure $p$ and we have shown that the WEC is fulfilled. We have presented the dynamical stability analysis under radial perturbations in terms of an effective potential. We have adopted a linear EoS of the form $p= \kappa \sigma $, the parameter $\kappa $ was subsequently restricted to the range $-1/2 < \kappa < 0$ in order to satisfy the requirements of our model. The matter at the shell can then be thought as some kind of dark energy. We have obtained that all physically possible configurations are dynamically stable.

We have found that the entropy of the extremally charged EBI shell is solely a function of the gravitational radius $r_{\ex}$, just as in the RN scenario \cite{Lemos2015a,Lemos2016}. This result is particularly interesting, since in the RN situation, the entropy only depends on $r_{\ex}$ due to the fact that $p=0$ for the extremally charged shell. However, in our case, despite the presence of a non-zero pressure, the dependence of the shell material mass $M$ on $r_{\ex}$ and the shell radius $R$ ensures that the overall entropy is only a function of $r_{\ex}$. Hence, just as in the RN case, extremally charged EBI thin shells sharing the same mass and charge, but of different radii, have the same entropy. This important result has a direct consequence when considering the thermodynamical stability of the shell: there exists only one non-trivial stability condition related to changes in $r_{\ex}$ alone, which in turn can be thought of changes in the shell charge $Q$.

A useful parametrization to display the stability regions in both the dynamical and the thermodynamical analyses is the ratio $b/Q$ between the nonlinearity parameter  $b$ and the extremal shell charge  $Q$, which has an upper and lower bound given by $0 \le b/Q < 2$. We have found that both the thermodynamical and dynamical stable regions grow as the nonlinear departures of the theory become evident. Further, the dynamically stable configurations do not provide any additional constraints on the radii of the shells and all physically viable shells are dynamically stable. On the other hand, the thermodynamical stability sets an upper bound on the value of the exponent $\omega$ in the power law adopted for the temperature equation of state, besides the lower bound ($\omega =-1$) required to avoid divergences in the temperature. Since the dynamical stability relates $\omega_\text{crit}$ with $\kappa$, completely stable configurations are those being thermodynamically stable with $-1< \omega \le \omega_\text{crit}$. This particular result is in accordance with the same analysis in a lower dimensional spacetime \cite{OlmosCayo2025}.

Several extensions of the present work are worth pursuing. A first natural step is to generalize our analysis to EBI thin shells outside the extremal scenario, where neither the mass $m$ nor the charge $Q$ are fixed by extremality, thereby enlarging the parameter space and testing whether a richer interplay between dynamical and thermodynamical stability is possible. It would also be interesting to revisit the thermodynamical stability problem by promoting the BI parameter $b$ to the status of a thermodynamic state variable, in line with various approaches to NLED black holes in which both the first law and the Smarr relation acquire additional contributions when this parameter is allowed to vary (c.f. \cite{Bokuliifmmodeacutecelsecfi2021,Gulin2017}). Finally, we note that in the limit $r \to 0$ the time–time component of the Einstein–Born–Infeld metric, though divergent, has the same qualitative functional form as that of Schwarzschild–AdS, which suggests interpreting $b^{-1}$ as an effective cosmological constant and formulating an extended \emph{thin–shell chemistry}, in analogy with the black hole chemistry program of Mann and collaborators \cite{Gunasekaran2012,Kubiznak2017}

\begin{acknowledgments}

This work was supported by CONICET (EFE and GFA) and by National Council for Scientific and Technological Development - CNPq and FAPERJ - Fundaçāo Carlos Chagas Filho de Amparo à Pesquisa do Estado do Rio de Janeiro, Processo SEI 260003/014960/2023 (MLP). We thank to the anonymous referee for helpful comments.

\end{acknowledgments}

\bibliography{exBIshell}

\end{document}